# Measurement of the optical dielectric function of transition metal dichalcogenide monolayers: $MoS_2$, $MoSe_2$, $WS_2$ and $WSe_2$

Yilei Li[1], Alexey Chernikov[1], Xian Zhang[2], Albert Rigosi[1], Heather M. Hill[1], Arend M. van der Zande[2,3], Daniel A. Chenet[2], En-Min Shih[1], James Hone[2], Tony F. Heinz[1*]

[1]*Departments of Physics and Electrical Engineering, Columbia University, 538 West 120th St., New York, NY 10027, USA*

[2]*Department of Mechanical Engineering, Columbia University, 500 West 120th Street, New York, NY 10027, USA*

[3]*Energy Frontier Research Center, Columbia University, 500 West 120th Street, New York, NY 10027, USA*

\* To whom correspondence should be addressed: tony.heinz@columbia.edu

Abstract: We report a determination of the complex in-plane dielectric function of monolayers of four transition metal dichalcogenides: $MoS_2$, $MoSe_2$, $WS_2$ and $WSe_2$, for photon energies from 1.5 – 3 eV. The results were obtained from reflection spectra using a Kramers-Kronig constrained variational analysis. From the dielectric functions, we obtain the absolute absorbance of the monolayers. We also provide a comparison of the dielectric function for the monolayers with the corresponding bulk materials.

## I. INTRODUCTION

Transition metal dichalcogenide (TMDC) crystals have emerged as a new class of semiconductors that display distinctive properties at monolayer thickness [1-3]. Their optical properties are of particular interest and importance. They exhibit a transition to direct band gap semiconductors at monolayer thickness [4,5], offer access to the valley degree of freedom by optical helicity [6-9], and display strong excitonic properties, with tightly bound neutral and charged excitons, as well as a non-hydrogenic Rydberg series of excited states [10-17]. The materials thus provide an excellent testing ground for the physics of 2D systems and many-body effects in solids. At the same time, they have attracted much interest for applications in optoelectronics as light emitters, detectors, and photovoltaic devices [18-25].

The most basic description of light-matter interactions in TMDC monolayers is given by the materials' complex dielectric function. The dielectric function provides a meeting point between experiment and theories of excited-state properties of this novel class of materials; knowledge of the dielectric function is also crucial for the characterization of these materials and for their use in emerging applications. Despite its central role, a systematic study of the optical dielectric functions for these materials has yet to be reported.

In this paper, we present a determination of the complex in-plane dielectric functions of four monolayer TMDCs – $MoS_2$, $MoSe_2$, $WS_2$, and $WSe_2$ – for the photon energies between 1.5 and 3 eV. Our study complements earlier ellipsometry measurements on monolayer $MoS_2$ [26-28]. We obtain the dielectric functions by Kramers-Kronig constrained analysis of the reflectance spectra of the monolayer samples supported on a transparent substrate. We also present calculated absorption spectra for the four different monolayers. The strong light-matter interaction leads to peaks in the imaginary part of the dielectric function for the A exciton exceeding 30 in some of the materials, with a corresponding single-layer absorption exceeding 15%. For the purpose of further understanding the resonance features in the dielectric response, we extract transition energies of resonances peaks and analyze several trends in the material response, including comparisons with previously reported bulk dielectric functions.

**II. OPTICAL REFLECTANCE MEASUREMENT**

We prepared the relevant monolayer TMDC samples on fused silica substrates by mechanical exfoliation of bulk crystals (2Dsemiconductors Inc.) [29] For comparison, we also examined selected monolayers prepared by chemical vapor deposition [30-34]. Samples of monolayer thickness were first identified by their optical contrast under a microscope with additional verification of thickness by Raman and photoluminescence spectroscopy.

The dielectric function was determined by reflectance measurements of the samples at room temperature. The reflectance measurements were performed using broadband emission from a tungsten halogen lamp, which was spatially filtered by a pinhole before being focused onto the sample using an objective. The spot size on sample was about 2 μm. The reflected light was collected by the same objective and deflected by a beam splitter to a spectrometer equipped with a CCD camera cooled to liquid-nitrogen temperature. We determined the reflectance spectra of the samples by normalizing the measured reflected power by that from the bare fused silica substrate. The reflectance from the fused silica substrate was calibrated using literature values for the material's refractive index [35]. The spectral resolution of our measurements was 2 meV, which was much narrower than any feature in the spectra. In our experimental configuration, the optical fields lie in the plane of the sample, thus only the *in-plane* dielectric response is accessed.

The absolute reflectance spectra for the TMDC monolayers on fused silica are presented in Figs. 1(a-d). For all four TMDC monolayers, the two lowest energy peaks in the reflectance spectra correspond to the excitonic features associated with interband transitions at the K (K') point in the Brillouin zone [36]. The two features, denoted by A and B, are attributed to the splitting of the valence band by spin-orbit coupling [37]. At higher photon energies we observe the spectrally broad response from higher-lying interband transitions [36], including the transitions near the Γ point [38,39].

The dielectric functions derived in this paper are obtained from exfoliated TMDC monolayer flakes. Different samples of the same material generally exhibit very similar optical response. This is illustrated by a comparison of the reflectance spectra for two different exfoliated samples of $MoS_2$ in Fig. 2(a). Charge doping, strain, and inhomogeneity can cause slight changes in the position and line widths of the narrow excitonic peaks in the spectral response. These differences are more apparent in comparisons with samples grown by chemical vapor deposition (CVD). Fig. 2(b) shows a comparison of the reflectance spectra for exfoliated and CVD-grown $MoS_2$ monolayers. Shifts in the A and B peaks of ~ 40 meV are observed, although the overall dielectric function is very similar.

## III. DETERMINATION OF THE COMPLEX DIELECTRIC FUNCTION BY KRAMERS-KRONIG CONSTRAINED ANALYSIS

We obtain the dielectric function $\varepsilon(E)$ for the four different TMDC monolayers from the experimental reflectance spectra by a Kramers-Kronig constrained analysis. We analyze the reflectance data treating the monolayer as a homogeneous medium with an effective thickness given by the interlayer spacing of the respective bulk material [36] ($d_{MoS2}$= 6.15 Å, $d_{MoSe2}$= 6.46

Å, $d_{WS2}$= 6.18 Å, and $d_{WSe2}$= 6.49 Å). The optical reflectance is calculated by the standard thin-film analysis [40], which fully accounts for all interference effects.

We model the complex dielectric function $\varepsilon(E) = \varepsilon_1(E) + i\varepsilon_2(E)$ of the samples as a function of photon energy $E$ using a superposition of Lorentzian oscillators:

$$\varepsilon(E) = 1 + \sum_{k=1}^{N} \frac{f_k}{E_k^2 - E^2 - iE\gamma_k} . \qquad (1)$$

Here $f_k$ and $\gamma_k$ are the oscillator strength and the line width of the $k$-th oscillator, and $E_k$ runs over the full spectral range. In our treatment, we choose a uniform spacing between the oscillators of 2 meV, much smaller than the narrowest feature in the optical spectra, and a full width of the individual oscillators of 10 meV to yield a spectrally smooth response. Note that the narrow width of the individual oscillators are only used for the modeling of an arbitrary dielectric function, and should not be associated with the lifetime of the electronic transitions. Satisfaction of the Kramers-Kronig relation is built into the functional form of Eq. (1) and does not need to be considered separately [41]. We fit the measured reflectance data using the dielectric function of Eq. (1), varying the oscillator strengths $f_k$ between 1.5 eV and 3 eV are varied to match the experimental data.

Our reflectance measurements only cover the spectral range of 1.5 eV $\leq E \leq$ 3 eV. Within our Kramers-Kronig constrained analysis, optical transitions outside this energy range need to be considered for an accurate determination of $\varepsilon$. We consider first the behavior at lower photon energies. In the infrared, the dielectric response arises from polar phonons and free carriers. Based on the weakness of the contribution of polar phonons in bulk materials [42,43] to the visible dielectric response, we can neglect this term. The influence of free carriers on the dielectric functions in the optical frequency range is estimated to be less than 0.1, using the Drude model for a carrier density of $10^{12}$ cm$^{-2}$, chosen on the basis of transport characteristics of usual TMDC transistor devices [4,44]. We therefore also neglect this contribution.

The influence of higher-energy electronic transitions, however, needs to be taken into account. Strong electronic transitions lie just beyond our measurement window. To address this issue, we make use of data for the bulk materials [45,46], including transitions up to a photon energy of $E = 30$ eV. As discussed below, the monolayer and bulk dielectric functions are quite similar at higher energies. We therefore expect the bulk dielectric function to provide an adequate approximation of the dielectric function to correct for the off-resonant response of higher-lying transitions.

With respect to experimental technique for the determination of the dielectric function, we note that rather than applying the Kramers-Kronig constrained analysis of the reflectance data, one could consider measurement of both reflection and transmission spectra of the sample. In principle, this yields two independent measurements for each photon energy, thus directly determining $\varepsilon_1$ and $\varepsilon_2$, without recourse to the Kramers-Kronig analysis. For a thin layer on a transparent substrate, as is the case here, however, both the transmission and reflection spectra are dominated by $\varepsilon_2$. Consequently, this approach cannot be applied reliably. We note that other researchers have recently applied ellipsometric techniques to determine the dielectric function of MoS$_2$ monolayers. [26-28]

Within the Kramer-Kronig constrained analysis, we find optimized dielectric functions that reproduce the measured reflectance spectra within the thickness of the lines in Fig. 1(a-d). In Fig. 1(e-l), we show the resultant real and imaginary parts of the dielectric functions for MoSe$_2$, WSe$_2$, MoS$_2$, and WS$_2$ over the spectral range of $1.5$ eV $\leq E \leq 3$ eV.

We can also express the material response in terms of the optical conductivity $\sigma(E) = \sigma_1(E) + i\sigma_2(E) = -i(\varepsilon_0 E/\hbar)[\varepsilon(E) - 1]$. We present this result in the form of the complex *sheet* conductivity $\sigma^s(E) \equiv \sigma(E)d$, where $d$ denotes the layer thickness. For a layer in which there is no significant propagation effect for light passing through the material, *i.e.*, $|\varepsilon(E)|^{\frac{1}{2}} (E/\hbar c)d \ll 1$, the sheet conductivity provides a full description of the optical response. In Figs. 3(a-d), we plot the real part of the sheet conductivity, $\sigma_1^s(E)$, for the four TMDCs.

In Figs. 3(e-h) we present the absorption spectra for free-standing monolayers of the four TMDC materials based on their measured dielectric functions. The overall absorbance (the fraction of the incident light absorbed by the material) in this frequency range is on the order of 10% for all four materials, demonstrating strong light-matter interaction even for a monolayer. In addition to these results for suspended monolayers, we have derived the absorbance for layers of each material on the fused silica substrate, based on the same dielectric function for the monolayers [Figs. 3 (i-l)]. The supported layers preserve the spectral shape for the absorbance for the suspended layers. In terms of the magnitude, however, the supported layers absorb about 1/3 less than the corresponding suspended layers. This can be understood from the local field correction factor of $4/(n_s+1)^2$ for the light intensity above the substrate, where $n_s \approx 1.46$ is the refractive index of fused silica.

## V. DISCUSSION

### VA. EXTRACTION OF THE A-B SPLITTING

To extract quantitative properties of the excitonic resonances, we parameterize $\varepsilon_2$ using a small, but physically meaningful number of Lorentzians terms, allowing for an extraction of transition energies of the excitonic features. [47] The resulting energy differences between the A and B excitonic transitions are shown in Fig. 4 for the different monolayers. Results for the four TMDC crystals, as well as for the exfoliated $MoTe_2$ monolayers [48], are presented as a function of the effective atomic number of the material, determined by weighting each element's relative contributions to the spin-orbit coupling [49].

The monolayer results correspond well with the previously reported data for $MoS_2$, $WS_2$, and $WSe_2$ monolayers [4,50]. The predicted A-B splitting from a three-band tight-binding model [51] is also shown in Fig. 4. The good agreement between our experiment and the predictions of the tight-binding model (fit to density functional calculations) is at first-sight surprising, since these calculations neglect many-body Coulomb effects. However, the correction to the transition energy from the (significant) exciton binding energy is largely offset by many-body corrections to the quasi-particle band gap, rendering the predicted transition energy within a single-particle calculation closer to experiment that might be expected. [52,53]

**VB. COMPARISON WITH THE DIELECTRIC FUNCTION OF BULK MATERIAL**

The dielectric functions for the monolayer TMDC crystals obtained in this work can be compared with the dielectric functions for the corresponding bulk materials [45,46], as shown in Fig. 5. While the dielectric functions for monolayer and bulk TMDCs show a significant overall similarity, differences in the spectral responses are readily seen. There is a broadening of the resonance features in the bulk materials comparing to the monolayers, which we attribute to the additional optical transitions and carrier relaxation channels arising from interlayer coupling. On the other hand, the total oscillator strength of the different features, which is proportional to the integrated area of the imaginary part of the dielectric function, is largely conserved.

We also observed that the resonance energies in the monolayer dielectric function are modestly shifted from the corresponding bulk material. In Fig. 6, we present the energy difference between the spectral peaks in the imaginary part of the dielectric functions for monolayers and the corresponding bulk material. We plot the A peak energy differences for $MoS_2$, $MoSe_2$, $WS_2$, and $WSe_2$, the C peaks energy differences for $MoS_2$, $MoSe_2$, and $WS_2$, and D peak energy difference for $WSe_2$. (The peak labels are shown in Fig. 5) We chose the D peak instead of the C peak for $WSe_2$ because the C peak in bulk $WSe_2$ overlaps the B peak, hindering an accurate determination of the peak energy. Comparing the monolayer to the bulk peak energies, we see blue shifts for the C/D peak of 150 – 300 meV, while the shifts for the A peak are in the range of 10 – 80 meV.

The qualitative trend of a shift of the peak energy to higher energies with decreasing thickness can be understood as a quantum-confinement effect. As mentioned above, the *net* many-body contribution to the transition energies is expected to be reduced significantly because of the cancellation of electron-electron and electron-hole interactions. Consequently, in considering the physical origin of the shift of the peaks from the bulk to the monolayer, we focus on quantum-confinement effects in the band structure. The A peak corresponds to the excitons located at the K-point in the Brillouin zone, which are composed primarily of the metal *d*-orbitals. Thus the A exciton is spatially localized in the plane of the metal atoms [54] and interlayer interactions in the bulk material do not lower the transition energy significantly. The C/D feature is, however, associated with transitions away from the K-point and involves a significant contribution from the chalcogen orbitals [38,49,54]. We therefore expect appreciably stronger interlayer interactions in building up the bulk material from monolayers. This provides a physical rationale for the larger observed red shift of the C/D transition compared to that of the A exciton in comparing the bulk material to the monolayer. The behavior of the C/D transition with increasing layer thickness is analogous to the quantum confinement effect of the lowest indirect optical transition[4,5], which involves states at Γ-point of the Brillouin zone [55].

## V. CONCLUSIONS

In conclusion, we have experimentally determined the complex in-plane dielectric functions of monolayer $MoS_2$, $MoSe_2$, $WS_2$ and $WSe_2$, over photon energies between 1.5 eV and 3.0 eV. The dielectric functions imply strong light-matter interactions even in monolayers of these TMDCs, with a peak absorbance of each of the four materials exceeding 15%. The A-B exciton splitting is extracted from the measured dielectric function and is found to be in good agreement with density functional calculations. Comparing to the bulk materials, we observed only a slight red shift of the A feature, but a more significant shift of the higher-lying (C/D) features. This behavior can be understood as a reflection of the different orbital character and interlayer interactions of the wavefunctions relevant for the lower and higher energy transitions.

The authors acknowledge the U.S. Department of Energy, Office of Basic Energy Sciences for support through Columbia Energy Frontier Research Center (grant DE-SC0001085) for the preparation and characterization of samples; the National Science Foundation through grants DMR-1122594 and DMR-1124894 for support for the optical measurements and data analysis. H. M. H. and A. R. were supported, respectively, by the NSF through an IGERT Fellowship (DGE-1069240) and by a Graduate Research Fellowship. A. C. acknowledges funding from the Alexander von Humboldt Foundation through a Feodor-Lynen Fellowship. The authors would like to thank Timothy C. Berkelbach, Tony Low, Dmytro Nykypanchuk, and Xiaoxiao Zhang for fruitful discussions.

**Figures:**

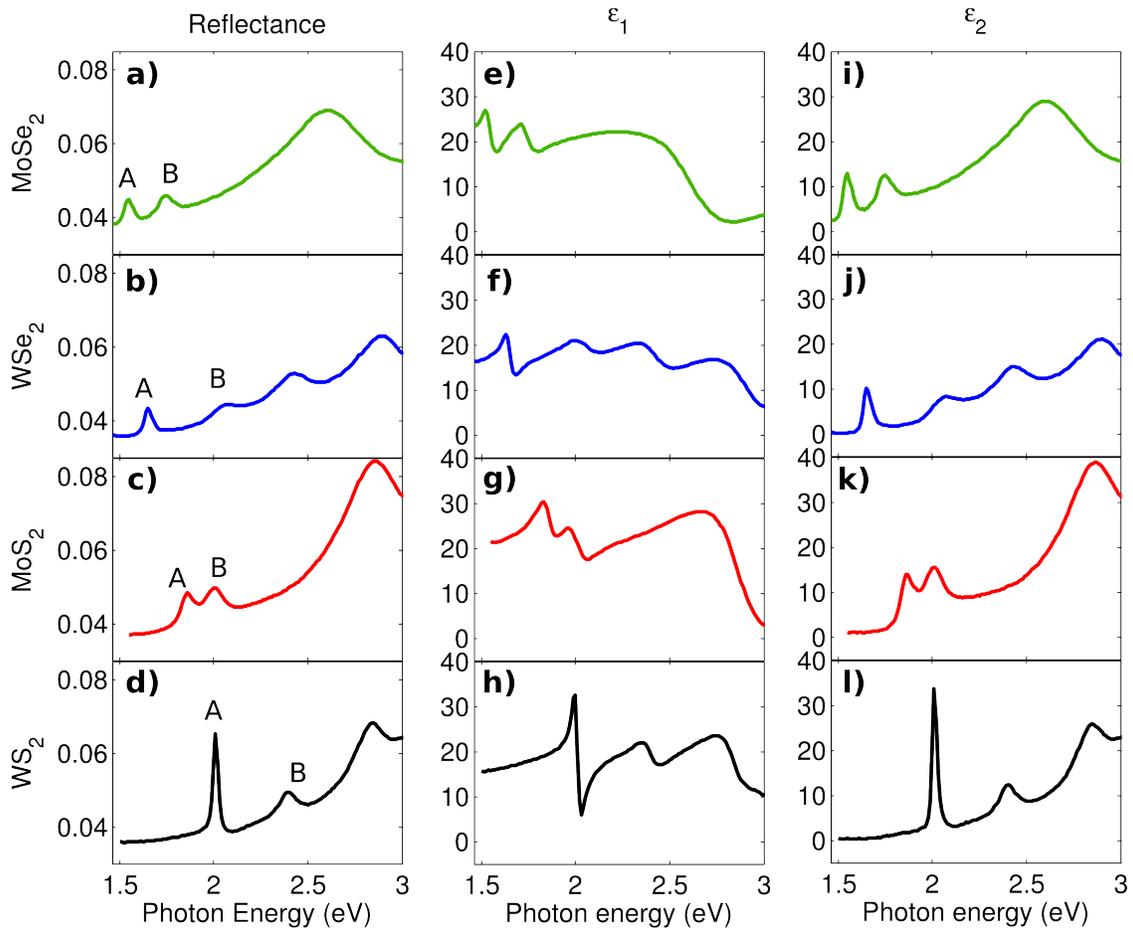

**Figure 1.** Optical response of monolayers of $MoSe_2$, $WSe_2$, $MoS_2$ and $WS_2$ exfoliated on fused silica: (a-d) Measured reflectance spectra. (e-h) Real part of the dielectric function, $\varepsilon_1$. (i-l) Imaginary part of the dielectric function, $\varepsilon_2$. The peaks labeled A and B in (a-d) correspond to excitons from the two spin-orbit split transitions at the K-point of the Brillouin zone.

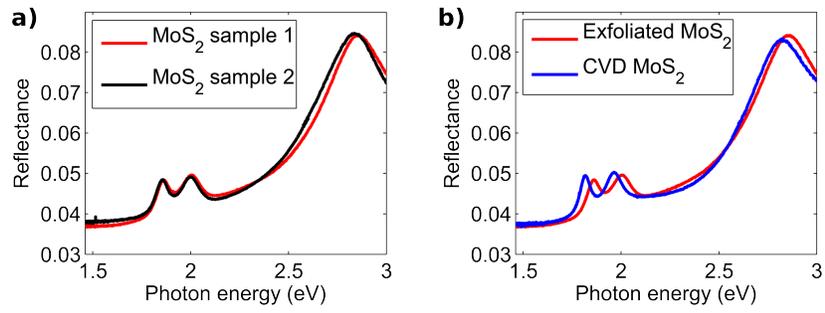

**Figure 2.** (a) Comparison of the reflectance spectra of two different exfoliated $MoS_2$ monolayers. (b) Comparison of the reflectance spectra of exfoliated (red) and CVD-grown (blue) $MoS_2$ monolayers.

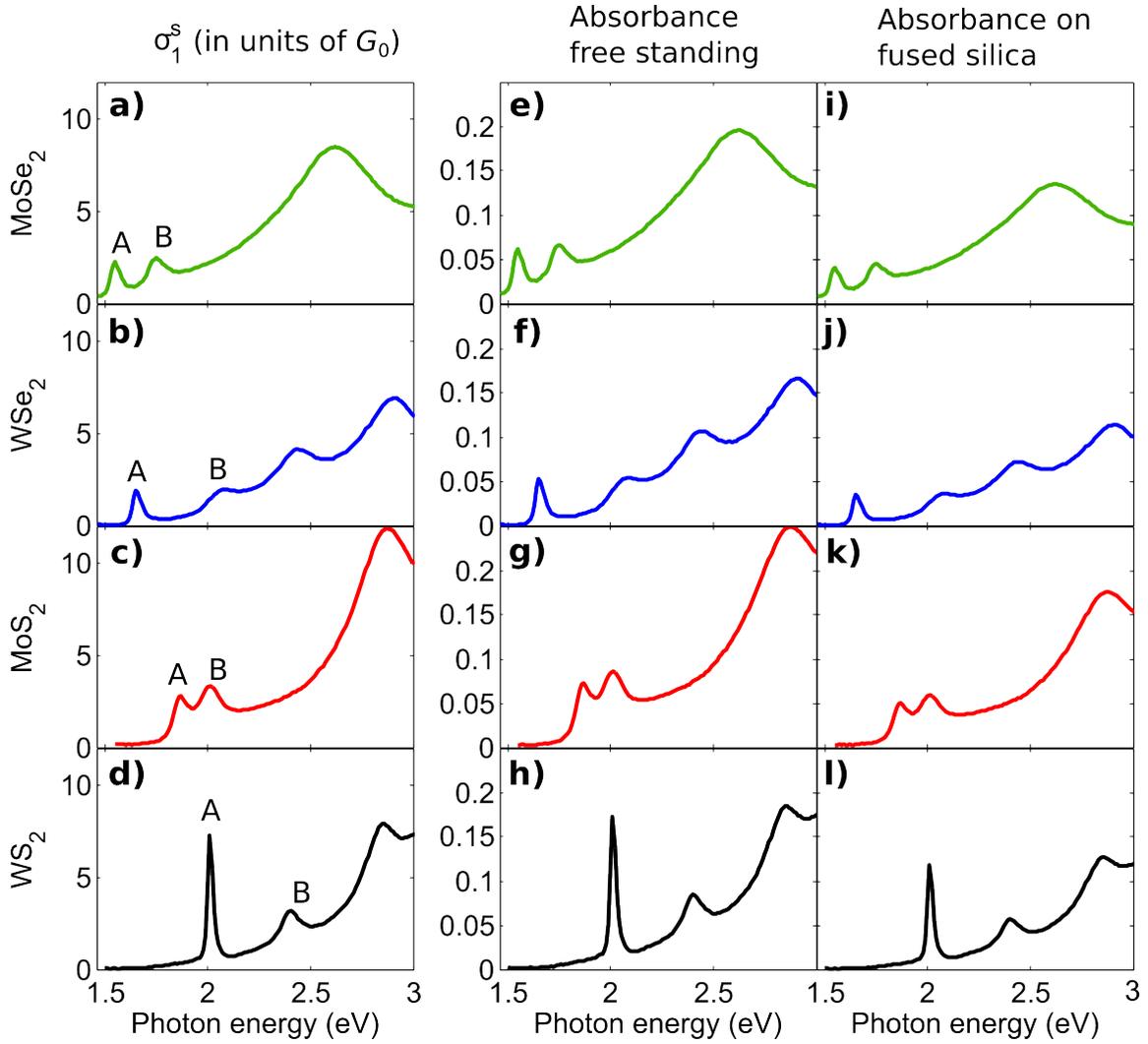

**Figure 3.** Real part of the sheet conductivity (in units of $G_0 = 2e^2/h$) and absorbance for (free-standing and supported) monolayers of MoSe$_2$ (a, e, i), WSe$_2$ (b, f, j), MoS$_2$ (c, g, k), and WS$_2$ (d, h, l). The peaks labeled A and B correspond to excitons from the two spin-orbit split transitions at the K-point of the Brillouin zone.

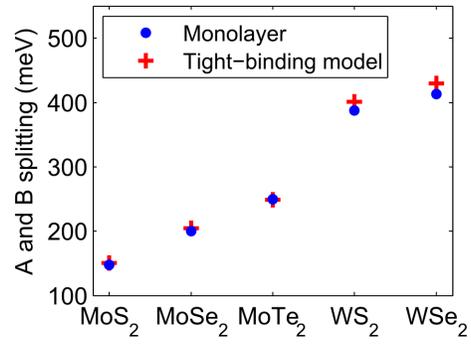

**Figure 4**. Energy difference between the A and B transitions (blue dots) in monolayer $MoSe_2$, $WSe_2$, $MoTe_2$, $MoS_2$ and $WS_2$. The result for $MoTe_2$ is from [48]. The predicted splittings (red crosses) within a three-band tight-binding model [51] are shown for comparison.

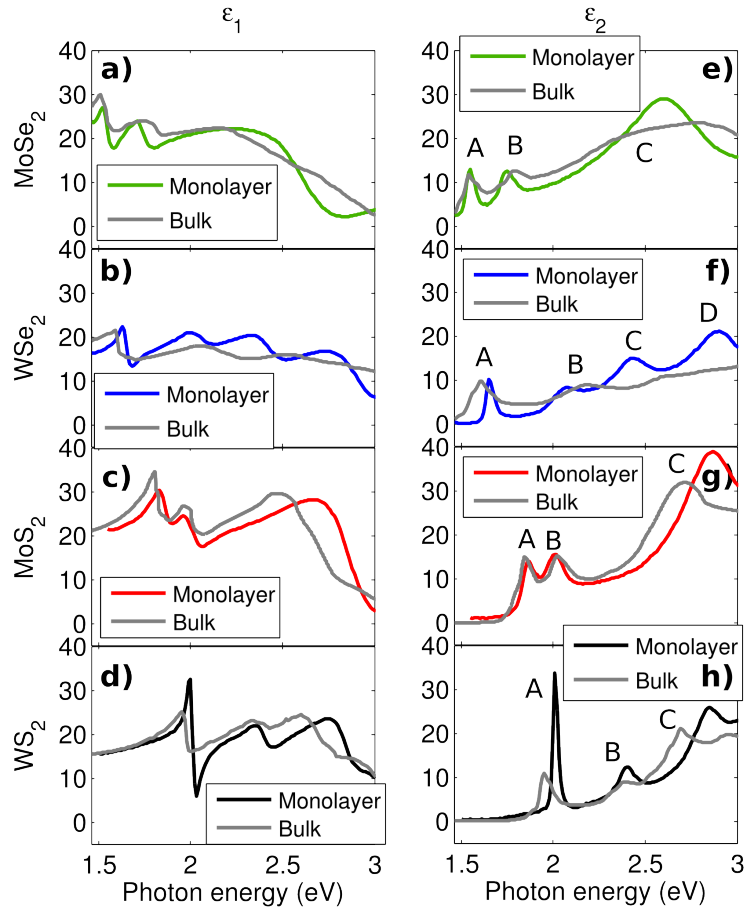

**Figure 5.** Comparison of the dielectric function of monolayer TMDC crystals (colored lines from Fig. 1) with that of the corresponding bulk material (gray) for $MoSe_2$ (a, e), $WSe_2$ (b, f), $MoS_2$ (c, g), and $WS_2$ (d, h). The peaks in $\varepsilon_2$ are labeled A, B, C and D. The energy differences of selected features are presented in Fig. 6.

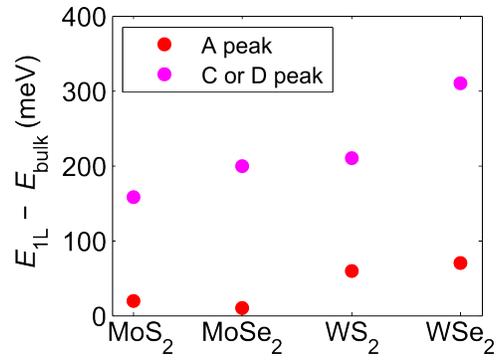

**Figure 6.** The observed blue shifts of features in the dielectric function of monolayer crystals of the TMDCs with respect to the corresponding bulk materials. The A peak denotes to the band-edge exciton at the K-point in the Brillouin zone and the C or D peak (depending on the material system, as discussed in the text) corresponds to the transitions away from the K-point.